\documentclass[aps,prd,superscriptaddress,twocolumn,showpacs,twoside,a4paper]{revtex4-1}
\usepackage{graphicx}
\usepackage{amsmath}
\usepackage{amsfonts}
\usepackage{bbm} 
\usepackage[utf8]{inputenc}
\usepackage{hyperref}
\usepackage{version}
\graphicspath{{Images/}}
\usepackage[titletoc]{appendix}

\usepackage{lipsum}
\usepackage{titlesec}

\titlespacing\section{0pt}{15pt plus 0pt minus 0pt}{6.3pt plus 0pt minus 0pt}

\newcommand\id{\ensuremath{\mathbbm{1}}} 
\DeclareMathOperator{\Tr}{Tr}

\begin{document}

\title{Renormalized entropy of entanglement in relativistic field theory}

\author{Issam Ibnouhsein} \email{issam.ibnouhsein@cea.fr} \affiliation{CEA-Saclay/IRFU/LARSIM, 91191 Gif-sur-Yvette, France} \affiliation{Université Paris-Sud, 91405 Orsay, France} 
\author{Fabio Costa} \affiliation{Faculty of Physics, University of Vienna, Boltzmanngasse 5, A-1090 Vienna, Austria}\affiliation{Institute of Quantum Optics and Quantum Information (IQOQI), Austrian Academy of Sciences, Boltzmanngasse 3, A-1090 Vienna, Austria}
\author{Alexei Grinbaum} \affiliation{CEA-Saclay/IRFU/LARSIM, 91191 Gif-sur-Yvette, France}

\begin{abstract}
Entanglement is defined between subsystems of a quantum system, and at fixed time two regions of space can be viewed as two subsystems of a relativistic quantum field. The entropy of entanglement between such subsystems is ill-defined unless an ultraviolet cutoff is introduced, but it still diverges in the continuum limit. This behavior is generic for arbitrary finite-energy states, hence a conceptual tension with the finite entanglement entropy typical of nonrelativistic quantum systems. We introduce a novel approach to explain the transition from infinite to finite entanglement, based on coarse graining the spatial resolution of the detectors measuring the field state. We show that states with a finite number of particles become localized, allowing an identification between a region of space and the nonrelativistic degrees of freedom of the particles therein contained, and that the renormalized entropy of finite-energy states reduces to the entanglement entropy of nonrelativistic quantum mechanics.
\end{abstract}


\maketitle

\section{INTRODUCTION}

Entanglement is a central concept in quantum theory and an important resource in quantum information theory. Different entanglement measures have been introduced to understand the structure of entanglement between finite-dimensional systems. One such measure is entanglement entropy, which can also be defined for pure states of extended quantum systems. Since all measurements on extended systems are performed within some finite region of space, it is natural in the relativistic context to study entanglement between localized subsystems, i.e. subsystems associated with spacelike separated regions of spacetime. Although natural, this is also problematic, because
relativistic field theories typically have infinite entanglement entropy between a region of space and its complement at fixed time \cite{bombelli_quantum_1986, srednicki_entropy_1993, callan_geometric_1994, calabrese_entanglement_2004,eisert_colloquium:_2010}. Indeed, the separation of the degrees of freedom of a quantum field into a finite region of space $A$ and its complement $\bar{A}$ implies a specific mapping associating $A$ and $\bar{A}$ with commuting subalgebras of the algebra of observables, called a localization scheme \cite{fleming_reeh-schlieder_1998,halvorson_reeh-schlieder_2001,piazza_quantum_2006,piazza_volumes_2007,piazza_glimmers_2010}. Different localization schemes correspond to different tensor product structures (TPS) of the total Hilbert space. Since entanglement is a property defined between subsystems, different localization schemes yield different values for entanglement between $A$ and $\bar{A}$ \cite{zanardi_virtual_2001,zanardi_quantum_2004}. A standard localization scheme associates to space regions the field operators and their conjugates therein defined. The vacuum state is entangled with respect to the TPS induced by the standard localization scheme, hence any two localized subsystems are correlated by modes at their boundary. Consequently, the entropy of entanglement between $A$ and $\bar{A}$ diverges due to the contribution of ultraviolet (UV) modes.

This peculiar behavior of entanglement is deeply rooted in the algebraic structure of relativistic quantum field theory (QFT). Indeed, it is a direct consequence of the Reeh-Schlieder theorem \cite{reeh_bemerkungen_1961} that all finite-energy states maximally violate Bell-like inequalities \cite{hegerfeldt_remark_1974,summers_vacuum_1985,summers_maximal_1987,redhead_more_1995,
summers_bells_1996,clifton_entanglement_2001, summers_yet_2008}. An operational content is given to this violation in \cite{keyl_infinitely_2002, keyl_entanglement_2006}: whenever the local algebras of observables of a bipartite system are not type I von Neumann algebras, maximally entangled states have infinite one-copy entanglement. For systems with type I local algebras of observables, states with infinite entropy of entanglement are trace-norm dense in state space \cite{eisert_quantification_2002}.

It follows from these arguments that entanglement measures for bipartite states with localized subsystems typically diverge at all energy scales when these are analyzed at a more fundamental level using QFT. Such subsystems do not yet correspond to the ones defined in the nonrelativistic regime, \emph{e.g.} qubits or trapped ions in quantum information protocols. The latter are described by finite dimensional Hilbert spaces and they generally produce finite results for entanglement measures. One expects that in the low-energy limit the degrees of freedom describing a region of space should simply correspond to the degrees of freedom of the particles therein contained. However, as a consequence of the Reeh-Schlieder theorem one cannot define local number operators, therefore finite-energy states cannot be localized \cite{fleming_reeh-schlieder_1998,halvorson_reeh-schlieder_2001}, making such a correspondence impossible. Hence a conceptual tension between the QFT description of entanglement for low-energy experiments and a description using nonrelativistic quantum theory.

In this paper, we introduce a novel approach to reconcile these two disagreeing notions of entanglement. Consider a scenario in which all possible field measurements are limited by some minimal spatial resolution $\epsilon$, thus restricting the algebra of observables to coarse-grained fields. The coarse-graining parameter $\epsilon$ has a clear operational meaning: since in practical situations it is impossible to resolve points in space with arbitrary precision, any realistic measurement of a field necessarily consists of a sample of a finite number of points, where each point corresponds to a finite region of space. We show that in the limit $\epsilon m \gg 1$, $m$ being the mass of a Klein-Gordon field, states with a finite number of particles become localized, allowing an identification between a region of space and the nonrelativistic degrees of freedom of the particles therein contained, and that the renormalized entropy of finite-energy states reduces to the one calculated in nonrelativistic quantum mechanics. This provides the missing controlled transition from the QFT picture of entanglement to entanglement in nonrelativistic quantum theory.

\section{ENTROPY OF ENTANGLEMENT IN QFT} Consider at fixed time a finite region of space $A$ and its complement $\bar{A}$. Region $A$ has two complementary descriptions: classical general relativity identifies it with a submanifold of Minkowski spacetime, but as a quantum subsystem, $A$ is described by a Hilbert space $\mathcal{H}(A)$, which is a factor in the tensor product decomposition $\mathcal{H}=\mathcal{H}(A)\otimes \mathcal{H}(\bar{A})$ of the total Hilbert space of the field theory under investigation. Suppose that the field is in a state $\rho$. The results of measurements to be performed in region $A$ are described by the reduced density matrix obtained by tracing out the degrees of freedom outside $A$: $\rho_A=\Tr_{\bar{A}} (\rho)$. The von Neumann entropy associated with region $A$ is then defined as $S_A =-\Tr (\rho_A \log \rho_A)$. This quantity typically requires some UV regulator in order to be well defined. Thus, in \cite{bombelli_quantum_1986}, a UV cutoff $\mu$ is introduced at the boundary between $A$ and $\bar{A}$, and in \cite{srednicki_entropy_1993,callan_geometric_1994,calabrese_entanglement_2004} a QFT is defined on a lattice of spacing $\mu$. References \cite{bombelli_quantum_1986, srednicki_entropy_1993, callan_geometric_1994, calabrese_entanglement_2004} show that for the vacuum state, the entropy of entanglement associated with region $A$ can be written as:
\begin{equation}
\label{old-results}
\frac{S_A}{\mathcal{A}}=C(\frac{\lambda}{\mu},m\mu)\mu^{-2},
\end{equation}
where $\mathcal{A}$ is the area of the boundary between $A$ and $\bar{A}$, $m$ the mass of the field, $\lambda$ an infrared cutoff and $C(x,y)$ some slowly varying function. For finite-energy states, power-law correction terms need to be added \cite{das_how_2006,das_power-law_2008}. These expressions diverge in the continuum limit for $m>0$ and, more generally, no cutoff-independent low-energy limit of the entropy can be derived using these approaches. 

A renormalization technique proposed in \cite{holzhey_geometric_1994} leads to states of negative entropy, which is not expected for a physically meaningful concept of state entropy.  Yet another approach consists in introducing a physical model of the measurement apparatus. One then derives an effective low-energy model of the measurement apparatus insensitive to vacuum entanglement of the underlying QFT \cite{costa_modeling_2009,zych_entanglement_2010}. However, this approach is strongly dependent on the choice of the model and does not provide a general and clear transition from the QFT picture of entanglement to entanglement in nonrelativistic quantum theory.

\section{COARSE-GRAINING PROCEDURE} For simplicity, we consider a neutral Klein-Gordon field of mass $m$ in one space dimension at fixed time (we put $\hbar = c = 1$). The algebra of local observables for the Klein-Gordon field is generated by the canonical field operators:
\begin{equation}
\label{fields}
\begin{aligned}
\hat \Phi(x)&=\int\frac{dk}{\sqrt{2\pi}} \frac{1}{\sqrt{2\omega_k}}\left(e^{i kx} \hat{a}_k+e^{-i kx}\hat{a}^{\dagger}_k\right),\\ 
\hat\Pi(x)&=-i\int \frac{dk}{\sqrt{2\pi}}\sqrt{\frac{\omega_k}{2}}\left(e^{i kx}\hat{a}_k-e^{-i kx}\hat{a}^{\dagger}_k\right),
\end{aligned}
\end{equation}
where $\omega_k=\sqrt{k^2+m^2}$ and $\hat{a}^\dagger_k$ creates a field excitation of momentum $k$.
The vacuum is defined by: 
\begin{equation}
\label{vacuum}
\hat{a}_k|\Omega\rangle = 0, \quad \forall k.
\end{equation}

\begin{figure}[!htbp]
\begin{center}
\includegraphics[scale=0.22]{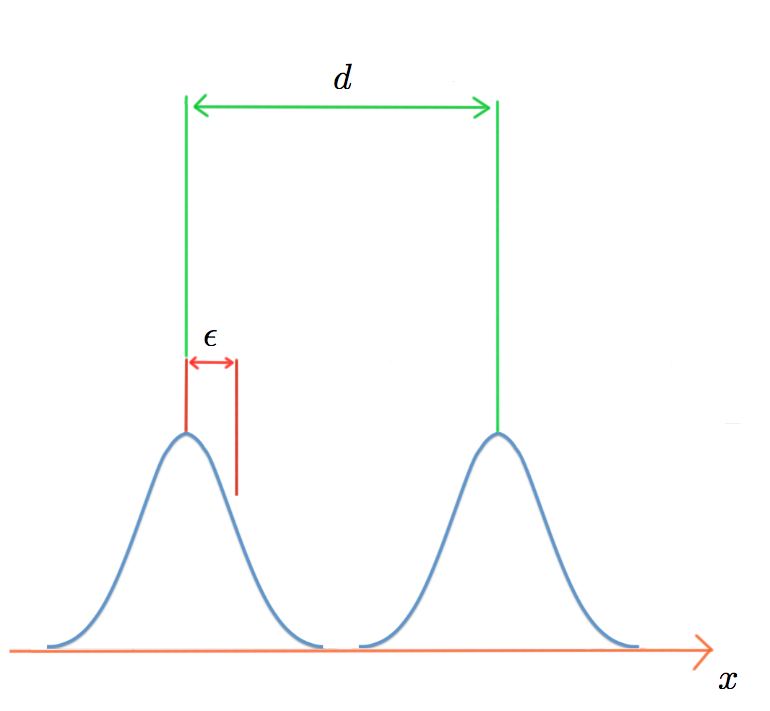}
\end{center}
\caption{The position in space at which a measurement is made can be determined only with limited accuracy, parametrized by $\epsilon$. This source of error is implemented by restricting the observable degrees of freedom to those accessible via measurement of coarse-grained operators. Neighboring profiles define different subsystems only if their separation $d$ verifies $d\gg\epsilon$. Under this condition, entanglement between neigboring profiles is a well-defined notion. We show that for finite-energy states, this entanglement reduces to the one calculated in nonrelativistic quantum theory.}
\label{schema}
\end{figure}

Assume that the resolution for distinguishing different points in space is bounded by some minimal length $\epsilon$. The algebra of observables that are accessible under such conditions is generated by the coarse-grained field operators (see Fig.\,\ref{schema}) : 
\begin{equation}
\label{smeared}
\begin{aligned}
\hat\Phi_{\epsilon}(x) &=  \int dy G_{\epsilon}(x-y) \hat\Phi(y), \\
\hat\Pi_{\epsilon}(x)  &= \int dy G_{\epsilon}(x-y) \hat\Pi(y).
\end{aligned}
\end{equation}
Function $G_{\epsilon}(x)$ describes the detection profile:
\begin{align}
		\label{eq:gauss}
		G_{\epsilon}(x) = \frac{1}{(2\pi \epsilon^2)^{1/4}} e^{- \frac{x^2}{4\epsilon^2}}.
\end{align}This choice of profile is natural if we interpret coarse graining as arising from a random error in the identification of a point in space. More generally, for any profile with a typical length $\epsilon'$, consider intervals of length $\epsilon$ on which the profile is approximately constant. One can then convolute such a profile with a Gaussian of variance $\epsilon$ and consider the limit $\epsilon m\rightarrow\infty$ instead of $\epsilon' m\rightarrow\infty$.

Define the operators: 
\begin{equation}
\label{effective}
\hat q_{j,\epsilon} =\hat\Phi_{\epsilon}(j d), \quad\hat p_{j,\epsilon} = \hat\Pi_{\epsilon}(j d),
\end{equation}where $d$ is the distance between neighbouring profiles. If $\epsilon\ll d$, they verify canonical commutation relations:
\begin{equation}
\label{CCR}
 \left[\hat{q}_{j,\epsilon},\,\hat{p}_{k,\epsilon}\right]\sim i\delta_{j\,k}.
 \end{equation}Imposing \eqref{CCR} is equivalent to saying that the operators $\{\hat{q}_{j,\epsilon},\hat{p}_{j,\epsilon}\}_j$ generate commuting subalgebras. Two commuting subalgebras of observables $\mathfrak{A}$ and $\mathfrak{B}$ that generate the whole algebra of observables induce a TPS on the Hilbert space of states: $\mathcal{H}=\mathcal{H}(A)\otimes \mathcal{H}(B)$ such that $\mathfrak{A} \rightarrow\mathfrak{A} \otimes \id_B$, $\mathfrak{B} \rightarrow \id_A \otimes \mathfrak{B}$ \cite{zanardi_virtual_2001,zanardi_quantum_2004}. The operators \eqref{effective} generate only a strict subalgebra of the entire algebra of field observables, because under coarse graining some possible observables are inaccessible. The whole algebra can be recovered by completing the set of functions $\{G_\epsilon(jd-y)\}_j$ up to an orthonormal basis in $\mathcal{L}^2(\mathbb{R})$ which, convoluted with the field operators \eqref{smeared}, defines a linear canonical transformation of modes. Thus, the algebra generated by the coarse-grained observables defines a decomposition of the total Hilbert space $\mathcal{H}={\cal H}_{cg}\otimes {\cal H}_{f}$, where ${\cal H}_{cg}$ are the coarse-grained, hence accessible, and ${\cal H}_{f}$ the fine-grained inaccessible degrees of freedom. The restriction to coarse-grained observables is therefore equivalent to tracing out subsystem ${\cal H}_{f}$, and operators $\{\hat{q}_{j, \epsilon},\hat{p}_{j, \epsilon}\}_j$ define distinct subsystems on $\mathcal{H}_{cg}$, each of which is isomorphic to a one-dimensional harmonic oscillator. Thus, we can define on $\mathcal{H}_{cg}$ the coarse-grained ladder operators: 
\begin{equation} 
\label{collectivecreation}
\begin{aligned}
\hat a_{j,\epsilon}= & \frac{1}{\sqrt{2}}\left(\sqrt{m'} \hat{q}_{j,\epsilon} + \frac{i}{\sqrt{m'}}\hat{p}_{j,\epsilon}\right),\\ 
 \hat a_{j,\epsilon}^\dag= & \frac{1}{\sqrt{2}}\left(\sqrt{m'} \hat{q}_{j,\epsilon} - \frac{i}{\sqrt{m'}}\hat{p}_{j,\epsilon}\right),
\end{aligned}
\end{equation}which verify $[\hat{a}_{j,\epsilon},\hat{a}_{k,\epsilon}^\dagger] \sim \delta_{jk}$. Parameter $m'$ has the dimension of mass. For a massive Klein-Gordon field, it is natural to take $m'=m$. Indeed, one can alternatively generate the local observables algebra with the ladder operators:
\begin{equation}
\begin{aligned}
\label{localcreation}
\hat{a}(x)&=\frac{1}{\sqrt{2}}\left(\sqrt{m}\hat\Phi(x)+\frac{i}{\sqrt{m}}\hat\Pi(x) \right),\\
 \hat{a}^\dagger(x)&=\frac{1}{\sqrt{2}}\left(\sqrt{m}\hat\Phi(x)-\frac{i}{\sqrt{m}}\hat\Pi(x)\right).
\end{aligned} 
\end{equation}Their coarse-grained versions correspond to operators in \eqref{collectivecreation} with $m'=m$. 

\section{THE NEWTON-WIGNER LOCALIZATION SCHEME} We recall that the Newton-Wigner (NW) annihilation and creation operators are respectively defined as the Fourier transforms of the momentum annihilation and creation operators \cite{newton_localized_1949}:
\begin{equation}
\label{NWcreation}
\begin{aligned} 
\hat a_{NW}(x) &=  \int \frac{dk}{\sqrt{2 \pi}} e^{i kx}\hat a_k ,\\
 \hat a_{NW}^{\dag}(x) &=  \int \frac{dk}{\sqrt{2 \pi}} e^{-i kx}\hat a^{\dag}_k.
\end{aligned}
\end{equation}The NW operators define a localization scheme \cite{fleming_reeh-schlieder_1998, halvorson_reeh-schlieder_2001,piazza_quantum_2006,piazza_volumes_2007,piazza_glimmers_2010} and are expressed in terms of the local fields \eqref{fields} as follows:
\begin{equation}
\label{NWfromfield}
\begin{aligned} 
\hat a_{NW}(x) =&  \frac{1}{\sqrt{2}}\int dy\left[R(x - y)\hat \Phi(y) +i R^{-1}(x - y)\hat \Pi(y)\right] ,\\ 
\hat a_{NW}^{\dag}(x) =&  \frac{1}{\sqrt{2}}\int dy\left[R(x - y)\hat \Phi(y) -i R^{-1}(x - y)\hat \Pi(y)\right],
\end{aligned}
\end{equation}
where we have introduced the functions:
\begin{equation}
\label{Rfunction}
R(x) = \int \frac{dk}{2\pi}\sqrt{\omega_k}e^{i kx} , \quad R^{-1}(x)= \int \frac{dk}{2\pi}\frac{1}{\sqrt{\omega_k}}e^{i kx}.
\end{equation}Operators $\{\hat a _{NW} (x)\}_x$ annihilate the global vacuum, therefore the global vacuum is a product state of local vacua. More generally, if local degrees of freedom are associated with the NW operators instead of the standard local fields \eqref{fields}, entropy of entanglement of the vacuum state is zero and entropy of finite-energy states, such as thermal states, becomes finite \cite{cacciatori_renormalized_2009}. However, identifying local degrees of freedom with NW operators at a fundamental level is problematic: the Hamiltonian of the field, expressed in terms of NW operators, is non-local. We do not address here the question of which localization scheme should be chosen at a fundamental level \cite{fleming_reeh-schlieder_1998,halvorson_reeh-schlieder_2001}. Instead we show that, under coarse graining, the entanglement properties of the NW fields for finite-energy states effectively hold, irrespective of the fundamental choice of local observables.

\section{CONVERGENCE BETWEEN LOCALIZATION SCHEMES} We now compare the algebra of coarse-grained observables generated by \eqref{collectivecreation} with the algebra generated by the following coarse-grained NW operators: 
\begin{equation} 
\label{smearedNW}
\begin{aligned}
\hat a_{NW,\epsilon}(x)  =& \int dy G_{\epsilon}(x-y)\hat a_{NW}(y),\\ 
\hat a_{NW,\epsilon}^\dag(x)  =& \int dy G_{\epsilon}(x-y) \hat a_{NW}^{\dag}(y) .
\end{aligned}
\end{equation}
Computations show that:
\begin{equation}
\label{smearedNWfromlocal}
\begin{split}
 \hat a_{NW,\epsilon}(x)  = \int dy \left[f^+_{\epsilon}(x-y)\hat a(y) + f^-_{\epsilon}(x-y) \hat a^{\dag}(y) \right],
\end{split}
\end{equation}
where:
\begin{equation}
\label{smearedR}
\begin{split}
	 f^{\pm}_{\epsilon}(x)=& \frac{1}{2} \left[\frac{R_{\epsilon}(x)}{\sqrt{m}} \pm \sqrt{m}R^{-1}_{\epsilon}(x) \right], \\
	 		   R_{\epsilon}(x)=& \int dy G_{\epsilon}(x-y)R(y) ,\\
    R^{-1}_{\epsilon}(x)=& \int dy G_{\epsilon}(x-y)R^{-1}(y).
\end{split}
\end{equation}In the limit of poor space resolution, the coarse-grained NW operators become indistinguishable from the coarse-grained local ladder operators since:
\begin{equation}
\label{fintegral}
\begin{aligned}
f^{\pm}_{\epsilon}(x) = & \frac{\sqrt{m}}{2} \int  \frac{dk}{\sqrt{2\pi}} e^{i mkx} G_{\frac{1}{2 m \epsilon}}(k)\\
&\cdot \left[	\left(1+k^2\right)^{1/4}\pm  	\left(1+k^2\right)^{-1/4} \right],
\end{aligned}
\end{equation}and in the limit where the minimal resolvable distances are much larger than the Compton wavelength, $\epsilon m  \gg 1$, the Gaussian $G_{\frac{1}{2 m \epsilon}}(k)$ verifies $G_{\frac{1}{2 m \epsilon}}(k)>0$ for $\left|k\right|\ll 1$ and $G_{\frac{2}{m \epsilon}}(k)\sim 0$ otherwise. Thus, in \eqref{fintegral} we have to integrate only over small values of $k$. We find:
\begin{equation}
\begin{aligned}
f^-_{\epsilon}(x) \sim & \, 0, \\
f^+_{\epsilon}(x) \sim & \, \sqrt{m} \int \frac{dk}{\sqrt{2\pi}} e^{i mkx} G_{\frac{1}{2 m \epsilon}}(k)=G_{\epsilon}(x).
\end{aligned}
\end{equation}
This result, plugged back into \eqref{smearedNWfromlocal}, gives:
\begin{equation}
\label{NWlocalconverge}
\hat a_{NW,\epsilon}(jd) \sim  \int dy G_{\epsilon}(jd-y) \hat a(y) = \hat a_{j,\epsilon}\quad \rm{for}\,\,\epsilon m  \gg 1 .
\end{equation}In the limit $\epsilon m \gg 1$, the coarse-grained NW operators still annihilate the global vacuum, hence the latter is a product state of effective local vacua. Equation \eqref{NWlocalconverge} then shows that the global vacuum is also a product state for the coarse-grained field operators. This implies that, in the limit of poor spatial resolution of detectors, an excitation localized ``around point $j$" is effectively described by applying the creation operator $\hat{a}^\dagger_{j,\epsilon}$ to the global vacuum $|\Omega\rangle$. Therefore, any one-particle state $|\psi\rangle = \int dk f(k) \hat{a}^{\dag}_k |\Omega\rangle$ can be effectively described as a sum $\sum_j \tilde{f}(jd) \hat a^{\dag}_{j,\epsilon} |\Omega\rangle$, where $f$ is a function verifying $\int dk |f(k)|^2 =1$ and $\tilde{f}$ its Fourier transform. As a consequence, such a state (which cannot be interpreted as localized in QFT unless it has infinite energy) can now be properly interpreted as localized, allowing a mapping between the description of a region of space in QFT and an effective description that only includes the nonrelativistic degrees of freedom therein contained. Hence, the structure of entanglement of any state with a finite number of excitations reduces to the entanglement between localized particles, i.e. to the standard, nonrelativistic, picture of entanglement. In particular, the entropy of entanglement of such states is upper bounded by the number of excitations times a factor describing how many states are available to each excitation (see the Appendix). Since finite-energy states correspond to states with a finite number of excitations, this result provides a controlled transition from the QFT picture of entanglement of finite-energy states to the nonrelativistic quantum theory one.

As an example, consider two mesons or two atoms with integer spin in a singlet state localized ``around points $i$ and $j$". In the QFT picture, entanglement between the region ``around point $i$" containing one meson with the rest of the system is infinite. Under the constraint of a bounded spatial resolution of detectors, the effective description of such a system in QFT is:
\begin{equation}
\begin{aligned}
&\frac{1}{\sqrt{2}}[\hat{a}^\dagger_{i, \epsilon,\uparrow}\hat{a}^\dagger_{j, \epsilon,\downarrow}-\hat{a}^\dagger_{i, \epsilon,\downarrow}\hat{a}^\dagger_{j, \epsilon,\uparrow}]|\Omega\rangle \\
=&\frac{1}{\sqrt{2}}\left(|0\rangle_1\cdots|0\rangle_{i-1}|\uparrow \rangle_i|0\rangle_{i+1}\cdots|0\rangle_{j-1} |\downarrow\rangle_j |0\rangle_{j+1}\cdots\right.\\
& \left.+\right|0\rangle_1\cdots|0\rangle_{i-1}|\downarrow \rangle_i|0\rangle_{i+1}\cdots|0\rangle_{j-1} |\uparrow\rangle_j |0\rangle_{j+1}\cdots),
\end{aligned}
\end{equation}which is formally equivalent to the state:
\begin{equation}
\frac{1}{\sqrt{2}}\left[|\uparrow\rangle_i |\downarrow\rangle_j - |\downarrow\rangle_i |\uparrow\rangle_j\right].
\end{equation}The entropy of entanglement between the region ``around $i$" and the rest of the system is then $S_i=\log(2)$, which is the expected value when modeling this system in nonrelativistic quantum theory. Note that by symmetry $S_i=S_j=\log(2)$.

\section{CONCLUSIONS} We have shown that in the limit of poor spatial resolution of the detectors, the entropy of entanglement of finite-energy states of a massive Klein-Gordon field reduces to the one calculated in nonrelativistic quantum theory. The derivation was independent of any effective low-energy model for the detectors. The results of this paper can be generalized to all noncritical bosonic systems, i.e. systems endowed with a finite length scale such as lattice models or models with local interactions and an energy gap (a natural length scale is then provided by the lattice spacing and the correlation length respectively) \cite{hastings_area_2007,masanes_area_2009, eisert_colloquium:_2010}. For critical systems, the correlation length diverges, hence different arguments are needed. For fermionic systems, an ambiguity in the definition of entanglement measures between subsystems arises due to the anticommutation of the creation and annihilation operators \cite{montero_fermionic_2011,friis_fermionic-mode_2013}. One can reformulate the problem of entanglement between localized subsystems in a purely algebraic way \cite{summers_vacuum_1985,summers_maximal_1987,yngvason_role_2005,matsui_spectral_2010,yngvason_localization_2014}, and a possible extension of the coarse-graining procedure to the algebraic framework is under investigation. 

\section*{ACKNOWLEDGMENTS} We thank \v{C}aslav Brukner, Magdalena Zych and Federico Piazza for helpful discussions. This work was supported by the European Commission Project Q-ESSENCE (No.\,248095), the European Commission Project RAQUEL, the John Templeton Foundation, FQXi, and the Austrian Science Fund (FWF) through CoQuS, SFB FoQuS, and the Individual Project No.\,2462.

\section*{APPENDIX: ENTANGLEMENT ENTROPY OF LOCALIZED SYSTEMS}
\label{Appendix}

Consider at fixed time a finite region of space $A$ and its complement $\bar{A}$. Suppose that the field is in a state with $N$ excitations. $A$ is decomposed into $M$ distinct regions $A_1,...,A_M$, whose points are assumed to be nonresolvable because of the limited spatial resolution of the detectors. An upper bound on the entropy of entanglement between subsystems $A$ and $\bar{A}$ is given by the dimension of the subspace of an $M$-mode system containing any number of particles between 0 and $N$:
\begin{equation}
D^M_N=\sum_{n=0}^N C^M_n = \frac{(M+N)!}{M!N!},
\end{equation}
where:
\begin{equation}
 C^M_n = \binom{M+n-1}{n} = \frac{(M+n-1)!}{n!(M-1)!}
\end{equation}is the dimension of the subspace with exactly $n$ particles. This provides an upper bound on the entropy of entanglement between $A$ and $\bar{A}$ for the $N$-particle state: 
\begin{equation}
\label{upper-bound-entropy-two-M-mode}
S_A \leq \log D^M_N.  
\end{equation}
If $M\gg N \geq 0$,  $\log D^M_N  \sim N\log M$, expressing the fact that the entropy of entanglement of such states is upper bounded by the number of excitations times a factor describing how many states are available to each excitation. One can encode degrees of freedom other than position by changing the value of $M$. For example, if two polarization states are available to each excitation, one must double the value of $M$.

\bibliographystyle{apsrev4-1}

\end{document}